\documentclass[conference]{IEEEtran}
\IEEEoverridecommandlockouts
\usepackage{graphicx} 
\usepackage{amsmath}
\usepackage{amsthm}
\usepackage{amssymb}
\usepackage{algorithm}
\usepackage{algpseudocode}
\usepackage{subcaption}
\usepackage{textcomp}
\usepackage{multirow}
\usepackage{xcolor}
\usepackage{booktabs}
\usepackage{tabularx}
\usepackage{siunitx}
\usepackage{ragged2e}  
\usepackage{cite}
\usepackage{geometry}
\newcolumntype{L}[1]{>{\hsize=#1\linewidth\raggedright\arraybackslash}X}
\usepackage{breqn}
\DeclareMathOperator*{\argmax}{argmax}
\DeclareMathOperator*{\argmin}{argmin}
\theoremstyle{definition}
\newtheorem{definition}{Definition}[section]
\graphicspath{ {images/} }
\algblock{Input}{EndInput}
\algnotext{EndInput}
\algblock{Output}{EndOutput}
\algnotext{EndOutput}
\algblock{Procedure}{EndProcedure}
\algnotext{EndProcedure}

\newtheorem{Remark}{Remark}
\newtheorem{Study}{Study}

\newcommand{\Desc}[2]{\State \makebox[2em][l]{#1}#2}

\begin{document}
\title{Nash or Stackelberg? -- A comparative study for game-theoretic AV decision-making
\thanks{This work is supported by Ford Motor Company.}
}

\author{\IEEEauthorblockN{Brady Bateman\IEEEauthorrefmark{1}}\thanks{\IEEEauthorrefmark{1}B. Bateman, M. Xin, and M. Liu are with the Department of Mechanical and Aerospace Engineering,
University of Missouri, Columbia, MO, USA (email: bgbg8p@umsystem.edu,  xin@missouri.edu, ml529@missouri.edu).
}
\and
\IEEEauthorblockN{Ming Xin\IEEEauthorrefmark{1}}
\and
\IEEEauthorblockN{H. Eric Tseng\IEEEauthorrefmark{2}}\thanks{\IEEEauthorrefmark{2}H. E. Tseng is with Ford Research and
Innovation Center, 2101 Village Road, Dearborn, MI 48124, USA (e-mail:
htseng@ford.com)}
\and
\IEEEauthorblockN{Mushuang Liu\IEEEauthorrefmark{1}}

}

\date{September 2023}

\maketitle

\begin{abstract}
This paper studies game-theoretic decision-making for autonomous vehicles (AVs). A receding horizon multi-player game is formulated to model the AV decision-making problem. Two classes of games, including Nash game and Stackelber games, are developed respectively. For each of the two games, two solution settings, including pairwise games and multi-player games, are introduced, respectively, to solve the game in multi-agent scenarios. Comparative studies are conducted via statistical simulations to gain understandings of the performance of the two classes of games and of the two solution settings, respectively. The simulations are conducted in intersection-crossing scenarios, and the game performance is quantified by three metrics: safety, travel efficiency, and computational time. 
\end{abstract}


\section{Introduction}

Autonomous vehicles (AVs) are expected to bring numerous social benefits, including reducing the number of crashes \cite{Wadud_MacKenzie_Leiby_2016}, improving mobility for people who may not be able to drive themselves \cite{Yang_Coughlin_2014}, and reducing traffic congestion \cite{Stanek_2017,Wang_Wang_Palacharla_Ikeuchi_2017}. However, technical challenges still remain to be addressed before fully autonomous driving is realized. The decision-making algorithm design is one of these challenges \cite{Gonzalez2016Review}, \cite{Kiran2022DRLSurvey}. Specifically, the AV decision-making is expected to generate safe and intelligent behaviors for the AVs when they interact with other road users, including human-driven vehicles, pedestrians, and bicycles. It is also expected that the algorithm is scalable to handle a large number of traffic agents and heterogeneous traffic scenarios. 

To model traffic agents' interactions, game-theoretic approaches have been explored in the literature \cite{9810195,jond2021autonomous,Chakeri_VehicleCroudsensing,lopez_lewis_liu_wan_nageshrao_filev_2022,liu_kolmanovsky_tseng_huang_filev_girard_2023,hang_lv_xing_huang_hu_2021,Ji2019StochasticStackelberg,Ji2021LaneMergingSF,Zhou_Xu_2021,Hang_IntegratedFramework,LiuThreeLevel,liu2022safe}. Game theory is a branch of mathematics that models the behaviors of reasonable players interacting with other reasonable players. Nash games \cite{9810195,jond2021autonomous,Chakeri_VehicleCroudsensing,lopez_lewis_liu_wan_nageshrao_filev_2022,liu_kolmanovsky_tseng_huang_filev_girard_2023,LiuThreeLevel,liu2022safe} are among the most widely-used games. Players in a Nash game make decisions simultaneously, and each player aims to optimize their self-interests while considering other players' possible decisions and corresponding interests. 
Such a game setting has been used in the context of autonomous driving. For example, in \cite{9810195}, an uncontrolled intersection scenario is considered and solved using Nash game. 
In \cite{jond2021autonomous}, a receding horizon Nash game is developed to control a convoy of autonomous vehicles. In \cite{Chakeri_VehicleCroudsensing}, researchers design an incentive mechanism modeled as a Nash game for a vehicle crowdsensing problem. 
Both \cite{jond2021autonomous} and \cite{Chakeri_VehicleCroudsensing} are limited in scope in that they only consider the scenarios where the surrounding vehicles are also AVs but not human-driven vehicles.
In \cite{lopez_lewis_liu_wan_nageshrao_filev_2022}, neural networks are implemented to create a Nash Q-learning algorithm. To ensure computational scalability, pairwise games, where the AV plays multiple 2-player Nash games with each of its neighbors, are employed \cite{lopez_lewis_liu_wan_nageshrao_filev_2022,LiuThreeLevel}. 
However, to use Nash games, care has to be taken to ensure the existence of a pure-strategy Nash equilibrium and the convergence of the solution-seeking algorithm, as specified by \cite{liu_kolmanovsky_tseng_huang_filev_girard_2023}. A special class of Nash game, called potential game is developed in \cite{liu_kolmanovsky_tseng_huang_filev_girard_2023,liu2022safe} to overcome challenges related to solving for the Nash equilibrium. 

Stackelberg games \cite{hang_lv_xing_huang_hu_2021,  Ji2019StochasticStackelberg, Ji2021LaneMergingSF, Zhou_Xu_2021, Hang_IntegratedFramework}, differing from Nash games by having leader/follower settings, have also been explored in the literature. A leader is a player who has the first-move advantage over the followers. Therefore, this setting is often used for scenarios where the road priorities are clear. However, in other scenarios, determining who the leader is not always straightforward. In \cite{Ji2021LaneMergingSF}, a Stackelberg game framework is utilized for lane-merging scenarios, in which the ego is assumed to be the leader and the vehicle behind the ego in the other lane is the follower. The target vehicles (i.e., the surrounding vehicles of interest) `politeness' is characterized in the cost function to allow the ego vehicle to gauge how successful it would be at changing lanes. 
In \cite{hang_lv_xing_huang_hu_2021}, Nash and Stackelberg games are separately combined with potential field methods and model predictive control in lane-changing and overtaking scenarios. According to this study, these two games exhibit different performance. Notably, the AVs maintain a larger gap between the AVs when each AV utilizes the Stackelberg equilibrium than when the AVs utilize the Nash equilibrium in a pairwise setting. 
However, \cite{hang_lv_xing_huang_hu_2021} tests their scenarios with only one situation, i.e., one set of initial conditions with one specific surrounding agents' behavior setting. 

Despite the above studies on game-theoretic autonomous driving, a deep understanding of when and how to select the appropriate games in a given scenario is still an open question. To answer this question, systematical analysis and comparison between various games, including Nash game and Stackelberg game, and various game settings, including multi-player and pairwise settings, are desired. This paper aims to bridge the gap and systematically investigate the AV performance (in terms \newgeometry{
 a4paper,
 left=54pt,
 top=54pt,
 bottom=104pt,
 right=37pt
 }of safety, travel efficiency, robustness to various surrounding vehicles' behavior, and algorithm computational cost) of these games.  

This paper is organized as follows. Section \ref{sec:ProbForm} formulates the AV decision-making problem as a receding horizon game problem in a generic setting. Section \ref{section:Nash} describes the Nash game. Section IV describes the Stackelberg game. Section V provides numerical studies, and Section VII concludes the paper.

\section{Problem Formulation}
\label{sec:ProbForm}
Consider a general traffic scenario, where $N$ traffic agents are sharing the road. 
Let $\mathcal{N}=\{1,2,...,N\}$ be the set of traffic agents. Each agent is subject to the discrete vehicle dynamics model in Eq. \eqref{eq1}.

\begin{equation}
 \label{eq1}
     x_i(t+1) = f_i(x_i(t),a_i(t)),
\end{equation}
where $x_i(t) \in \mathcal{X}_i$ is the state of agent $i$ at time $t$, $a_i(t) \in \mathcal{A}_i$ is the action of agent $i$ at time $t$. 
Let $\mathbb{A}$ be the Cartesian product of $\mathcal{A}_i$, i.e., $\mathbb{A}=\mathcal{A}_1 \times \mathcal{A}_2 \times ... \mathcal{A}_N$, and $(a_i(t),a_{-i}(t)) \in \mathbb{A}$ represent all agents' actions at time $t$, where $-i$ denotes all other agents except agent $i$, i.e $-i=\{1,2,...,i-1,i+1,...,N\}$. Define $x(t)=\{x_i(t),x_{-i}(t)\}$ as the global system state, $a(t)=\{a_i(t),a_{-i}(t)\}$ as the global action. 

Each agent has its own objectives in a driving scenario, such as tracking its desired speeds.  The performance index of agent $i$ at time step $t$ is denoted as $Q_i(x(t),a(t))$. At time $t$, each agent aims to find its action sequence (also called strategies) over $T$ $ (T\geq1)$ time steps, i.e., $s_i(t)=\{a_i(t),a_i(t+1),...,a_i(t+T-1) \}$ to optimize its cumulative performance. 
\begin{equation}
    s_i^*(t) \in \argmin_{s_i(t) \in S_i} J_i(x(t),s(t)),
    \label{eq:problem}
\end{equation}
\begin{equation}
   J_i(x(t),s(t)) = \sum_{\tau=t}^{t+T-1} Q_i(x(\tau),a(\tau)),
   \label{eq:cumulativePerformance}
\end{equation}
where $J_i:\mathbb{S}\times\mathcal{X}\mapsto\mathbb{R}$, $S_i$ is the set of all possible strategies available to agent $i$, and $\mathbb{S} = \mathcal{S}_1 \times \mathcal{S}_2 \times ... \mathcal{S}_N$ be the strategy space. Let $s(t)=\{s_i(t),s_{-i}(t)\}\in \mathbb{S}$ be the global strategy, or called strategy profile. For this paper, we only consider games with finite strategy spaces.  

The challenge to this multi-agent decision-making problem is that each agent has their own driving performance, which is mutually affected by other road users' strategies, as shown in Eq. \eqref{eq:problem}-\eqref{eq:cumulativePerformance}. To handle such a coupled optimization, we investigate game-theoretic approaches. We denote the game described in Eq. \eqref{eq:problem}-\eqref{eq:cumulativePerformance} as $ \mathcal{G} = \{\mathcal{N},\mathbb{S},\{J_i\}_{i \in \mathcal{N}}\}$. For clarity, superscripts and subscripts may be used to signify which type of game is being played. The superscripts N and S shall refer to Nash and Stackelberg games, respectively. A subscript $k$ shall denote a pairwise game between players $1,k\in\mathcal{N}$, e.g. $\mathcal{G}^N_2 = \{\{1,2\},\mathcal{S}_1 \times \mathcal{S}_2,\{J_i\}_{i \in \{1,2\}}\}$ refers to a Nash game played between players $1$ and $2$, which are players in $\mathcal{G}^N$. We assume the ego vehicle is the player 1 in all games. The detailed game settings employed in this paper are described in Section \ref{section:Nash}-\ref{section:Stackelberg}. 

\section{Nash Game}
\label{section:Nash}
In a Nash game, players are considered to be symmetric, i.e., no player has special advantages over others. In a Nash game, each player aims to use their best response to other players' strategies. The best response is defined as follows:
\begin{definition}[Best Response \cite{lã_chew_soong_2016}]
A strategy $s_i^*(t)$ is a best response to other players' fixed strategies $s_{-i}(t)$ if and only if
\begin{equation*}
    J_i(x(t),\{s_i^*(t),s_{-i}(t)\}) \leq J_i(x(t),\{s_i(t),s_{-i}(t)\}), \forall s_i \in S_i.
\end{equation*}
\end{definition}

If everyone plays their best response, then no player can get a lower cost by only changing their strategies. This is the Nash equilibrium and is defined as below:
\begin{definition}[Nash Equilibrium \cite{lã_chew_soong_2016}]
    A global strategy $\{s_i^*(t),s_{-i}^*(t)\}$ is a Nash equilibrium if and only if
    \begin{dmath*}
        J_i(x(t),\{s_i^*(t),s_{-i}^*(t)\}) \leq J_i(x(t),\{s_i(t),s_{-i}^*(t)\}), {\forall i, \forall s_i \in S_i.}
    \end{dmath*}
\end{definition}
Nash equilibrium is among the most widely-used solution concepts in a Nash game.  
To seek for Nash equilibrium, the best response dynamics algorithm \cite{Durand_Gaujal_2016} is often utilized, where, starting from an initial strategy profile, each player's best response is iteratively solved until convergence. Please see Algorithm \ref{alg:NEBRD} for the detailed procedures. 
Despite its wide acceptance, a Nash game also has limitations. Specifically,
\begin{enumerate}
    \item Given a general multi-player game as in Eq. \eqref{eq:problem}, there is no guarantee that a Nash equilibrium always exists.
    \item Given a general multi-player game,  even if a Nash equilibrium exists, there is also no guarantee that algorithm \ref{alg:NEBRD} converges.
    \item Algorithm \ref{alg:NEBRD} is computationally demanding. An $N$-player game with $|\mathbb{S}|$ possible strategies, where $|\mathbb{S}|$ represents the cardinality of $\mathbb{S}$, can have  $O(N|\mathbb{S}|^{N-1})$ computational complexity \cite{Durand_Gaujal_2016}, making it unpractical to solve in real time.
\end{enumerate}

To solve the computational challenge, a pairwise game setting has been used in the literature \cite{hang_lv_xing_huang_hu_2021}. In the pairwise game, a $2$-player game for every pair of agents that includes the ego vehicle is solved. After solving the $N-1$ $2$-player games, the most conservative ego vehicle decision is then selected as the final decision for the ego vehicle. The detailed procedures of the pairwise games are described in Algorithm \ref{alg:NEBRD_pairs}, where $\text{MC}(S^*_{ego}(t))$ means the most conservative strategy in the strategy set  $S^*_{ego}(t)$. 

\begin{algorithm}[ht]

\caption{Nash Equilibrium Best Response Dynamics}
\label{alg:NEBRD}
\begin{algorithmic}
 
 \Input
    \Desc{$\mathcal{G}^N$}\hspace*{0.2in}{Nash Game to solve}
  
    \EndInput
    \Output
    \Desc{$s^{*,N}(t)$}\hspace*{0.2in}{Nash Equilibrium}
    \EndOutput
    \State {\bf Procedure}
    \end{algorithmic}
    \begin{algorithmic}[1]
    \setcounter{ALG@line}{0}

\State \hspace*{0.2in} Initialize $s^{*,N}(t)\in\mathbb{S}$
\State \hspace*{0.2in} NECondition = 0
\State \hspace*{0.2in} {\bf While} {NECondiiton=0} {\bf do}
\State \hspace*{0.3in}$s_{prev}(t) = s^{*,N}(t)$
\State \hspace*{0.3in}{\bf For each} {$i \in \mathcal{N}$} {\bf do}
\State \hspace*{0.4in}$s_i^*(t) = \argmin_{s_i \in S_i} J_{i}(x(t),\{s_i(t),s_{-i}^*(t)\})$
\State \hspace*{0.3in}{\bf  End For}
\State \hspace*{0.3in}$s^{*,N}(t) = \{s_1^*(t),s_2^*(t),...,s_N^*(t)\}$
\State \hspace*{0.3in} {\bf If} {$s^{*,N}(t)=s_{prev}(t)$} {\bf then}
\State \hspace*{0.4in} NECondition = 1
\State \hspace*{0.3in}{\bf  End IF}
\State \hspace*{0.2in}{\bf  End While}

 \end{algorithmic}
 
\end{algorithm}

\begin{algorithm}[ht]

\caption{Nash Equilibrium Pairwise Best Response Dynamics}
\label{alg:NEBRD_pairs}
\begin{algorithmic}
 
 \Input
    \Desc{$\mathcal{G}^N$}\hspace*{0.2in}{Nash Game to solve}
  
    \EndInput
    \Output
    \Desc{$s^{*,Np}$(t)}\hspace*{0.2in}{Nash pairwise equilibrium}
    
    \EndOutput
    \State {\bf Procedure}
    \end{algorithmic}
    \begin{algorithmic}[1]
    \setcounter{ALG@line}{0}
    
   \State \hspace*{0.2in} {$S^*_{ego}(t)=\{\}$}
\State \hspace*{0.2in}{\bf For} {$i = 2:N$} {\bf do}
\State \hspace*{0.3in}{$\mathcal{G}_i^N=\{\{1,i\},\mathcal{S}_1 \times \mathcal{S}_i,\{J_i\}_{i \in \{1,i\}}\}$}
\State \hspace*{0.3in} {$[s^*_{ego}(t),s_{i}^*(t)]=$Algorithm1($\mathcal{G}_i^N$)}
\State \hspace*{0.3in} {$S^*_{ego}(t)=S^*_{ego}(t)\cup \{s^*_{ego}(t)\}$}
\State \hspace*{0.2in}{\bf End For}
\State \hspace*{0.3in} {$s^*_1(t)=\text{MC}(S^*_{ego}(t))$}
\State \hspace*{0.3in} {$s^{*,Np}(t) = \{s_1^*(t),s_2^*(t),...,s_N^*(t)\}$}
 \end{algorithmic}
 
\end{algorithm}

\section{Stackelberg Games}
\label{section:Stackelberg}

In a Stackelberg game, the players are not treated symmetrically. Instead, certain player(s) (called leader(s)) are designed to have first-move advantages over others (called followers). A leader first selects its strategy to optimize its self-interest. The followers then play their best response to the leader's strategy\cite{Korzhyk_Yin_Kiekintveld_Conitzer_Tambe_2011},\cite{Simaan_Cruz_1973}. While everyone still aims to optimize their self-interests,  this hierarchical structure may lead to different behaviors from the Nash game. If a follower has multiple best responses to a given strategy of the leader, the follower can choose whether to be in favor of the leader or the opposite.  This motivates two outcomes, called strong and weak Stackelberg equilibrium, respectively.
\begin{definition}[Strong Stackelberg Equilibrium \cite{guo2018inducibility}] The strategy set $\{s_L^*(t),s_F^*(t)\}$, where L and F refer to leader and follower, is a strong Stackelberg equilibrium if and only if
 \begin{equation*}
     \{s_L^*(t),s_F^*(t)\} = \argmin_{s_L \in S_L} \min_{s_F \in BR(s_L)}{J_L(x(t),\{s_L(t),s_F(t)\})}, 
 \end{equation*}
 where $BR(s_L)$ is the set of the followers' best responses to the leader's strategy $s_L(t)$.
 \label{def: SSE}
 \end{definition}
 \begin{definition}[Weak Stackelberg Equilibrium \cite{guo2018inducibility}] The strategy set $\{s_L^*(t),s_F^*(t)\}$, where L and F refer to leader and follower, is a weak Stackelberg equilibrium if and only if
 \begin{equation*}
     \{s_L^*(t),s_F^*(t)\} = \argmin_{s_L \in S_L} \max_{s_F \in BR(s_L)}{J_L(x(t),\{s_L(t),s_F(t)\})}, 
 \end{equation*}
 where $BR(s_L)$ is the set of the followers' best responses to the leader's strategy $s_L(t)$.
 \label{def: WSE}
 \end{definition}
Many studies in the literature \cite{Wang2020DecisionReview}, \cite{Ji2019StochasticStackelberg}, \cite{Flad2017}, \cite{DBLP:journals/corr/abs-2009-06394} do not make a distinction between these two Stackelberg equilibria in the sense that they do not consider the cases when a follower has multiple best responses to the leader. Few studies specify the Stackelberg equilibrium they employ (for example,  \cite{hang_lv_xing_huang_hu_2021} uses the strong, and \cite{Ji2021LaneMergingSF} uses the weak), however, without analysis on how and why to select the appropriate Stackelberg equilibrium given a driving scenario. 
We here include a brief discussion on the strong and weak Stackelberg equilibria selection in Remark \ref{remark:SvW}, and a comparative study between the two is performed in Study \ref{Study:TwoStacks}. 
 
Algorithm \ref{alg:Strong Stackelberg Equilibrium} shows the procedures to determine the strong and weak Stackelberg equilibria in a $2$-player game. In this algorithm, the followers' best responses to each of the leaders' possible strategies are calculated first. With this, the leader's optimal strategy corresponding to its minimum cost is then selected. 
Given the leader's strategy, the follower then chooses their strategies to optimize their self-interests. As we can see from these procedures, the leader in a Stackelberg game has the privilege over the follower, thanks to its first-move advantage. Note that the strong and weak Stackelberg equilibria are similar enough that we use a conditional statement to denote the difference between the two.


\begin{algorithm}[ht]
    \caption{Stackelberg Equilibrium - Two Players}
    \label{alg:Strong Stackelberg Equilibrium}
    \begin{algorithmic}
    \Input
   \Desc{$\mathcal{G}^S$}\hspace*{0.2in}{Stackelberg Game to solve}
    \EndInput
    
    \Output
    \Desc{$s_L^*(t)$}\hspace*{0.2in}{Optimal leader strategy}
    \Desc{$s_F^*(t)$}\hspace*{0.2in}{Optimal follower strategy}
    \EndOutput
    \State {\bf Procedure}
    \end{algorithmic}
    \begin{algorithmic}[1]
    
    \State \hspace*{0.2in}{\bf For each} $s_L \in S_L$ {\bf do}
    \State \hspace*{0.24in}$BR(s_L) = \argmin_{s_F \in S_F} J_F(x(t),\{s_L(t),s_F(t)\})$
   \State \hspace*{0.3in}{\bf If} {playing strong Stackelberg game}
    \State \hspace*{0.3in}$s_F^{*s_l} = \argmin_{s_F \in BR(s_L)} J_L(x(t),\{s_L(t),s_F(t)\}) $ 
    \State \hspace*{0.3in}{\bf Else If} {playing weak Stackelberg game}
      \State \hspace*{0.3in}$s_F^{*s_l} = \argmax_{s_F \in BR(s_L)} J_L(x(t),\{s_L(t),s_F(t)\}) $ 
    \State \hspace*{0.3in}{\bf End If}
    \State \hspace*{0.2in}{\bf End For}
    \State \hspace*{0.24in}$s_L^*(t)= \argmin_{s_L \in S_L} J_L(x(t),\{s_L(t),s_F^{*s_l}(t)\})$
    \State \hspace*{0.24in}$s_F^*(t)= \argmin_{s_F \in S_F} J_F(x(t),\{s_L^*(t),s_F(t)\})$
    
    \end{algorithmic}
\end{algorithm}

To deal with multiple players,  two mechanisms, called hierarchical game \cite{Conitzer_Sandholm_2006} and pairwise game \cite{hang_lv_xing_huang_hu_2021}, have been developed in the literature.  We detail the corresponding algorithms in Algorithms \ref{alg:Hierarchy Stackelberg Equilibrium} and \ref{alg:SE_pairs}, respectively.

In a hierarchical game setting, a player $i$ plays its best response to the players higher in the hierarchy (thus leaders to player $i$), while considering the objectives of players lower in the  hierarchy (thus followers to player $i$). Such a hierarchical Stackelberg game algorithm is studied in \cite{Conitzer_Sandholm_2006}, but has not been widely used in the context of autonomous driving due to its high complexity to solve \cite{Wang2020DecisionReview,Li_Yao_Kolmanovsky_Atkins_Girard_2022}. 

Algorithm \ref{alg:Hierarchy Stackelberg Equilibrium} is defined recursively. It makes use of subgames played between one leader and one follower. Let $\mathcal{G}^S$ be the game to solve. The function $\text{Order}(\mathcal{N})$ on line \ref{algline:order} of algorithm \ref{alg:Hierarchy Stackelberg Equilibrium} takes the set of players $\mathcal{N}$ and determines who the leader $L$ and immediate follower $F$ of the game are. Suppose the leader and follower of the game are the players $i,j$, respectively. Player $i$ uses player $j$'s best responses to inform it's optimal strategy according to Definition \ref{def: SSE} or \ref{def: WSE}. But player $j$ may be a leader to another player $k$. To determine player $j$'s best response, a subgame $\mathcal{G}_{sub}$, which is the game that results after player $i$ selects their strategy $s_L$, is played between leader $j$ and follower $k$.  How one determines this order is generally dependent on the specifics of the scenario. We use standard road priority rules to determine this order, similarly to \cite{Li_Yao_Kolmanovsky_Atkins_Girard_2022}. Also, we denote a set of followers who have selected a strategy in the hierarchical game as $F+\subset\mathcal{N}$, such that $s_{F+}^*(t)$ is the optimal strategies of players $F+$ at time $t$. The pairwise game setting is similar to the ones as in Algorithm \ref{alg:NEBRD_pairs}, and most Stackelberg game studies \cite{hang_lv_xing_huang_hu_2021,Ji2021LaneMergingSF,Li_Yao_Kolmanovsky_Atkins_Girard_2022,Liu_Li_Tseng_Kolmanovsky_Girard_2023} use this setting.

\begin{algorithm}[ht]
    \caption{Stackelberg Equilibrium: Hierarchy \cite{Conitzer_Sandholm_2006}}
    \label{alg:Hierarchy Stackelberg Equilibrium}
    \begin{algorithmic}
    \Input
    \Desc{$\mathcal{G}^S$}\hspace*{0.2in}{Stackelberg game to solve}
    \EndInput
    \Output
    \Desc{$s^{*,Sh}(t)$}\hspace*{0.2in}{Stackelberg hierarchy equilibrium}
    \EndOutput
    \State {\bf Procedure}
    \end{algorithmic}
    \begin{algorithmic}[1]

     \State \hspace*{0.2in} $[L,F] = \text{Order}(\mathcal{N})$\label{algline:order}
    \newline\Comment Determine hierarchy based on road priority rules
    
    \State \hspace*{0.2in}{\bf For each} $s_L \in S_L$ {\bf do}
    \State \hspace*{0.3in}$S_L^*=\{s_L\}$
    \State \hspace*{0.3in}$\mathbb{S}^*=S_L^* \times S_F$
    \newline\Comment{Fix the current leaders strategy, then solve a subgame for the next player in the hierarchy}
    \State \hspace*{0.3in}$\mathcal{G}_{sub}=\{\mathcal{N},\mathbb{S}^*,\{J_i\}_{i \in \mathcal{N}}\}$
    \State \hspace*{0.3in}$BR(s_L) = \text{Algorithm \ref{alg:Hierarchy Stackelberg Equilibrium}} (\mathcal{G}_{sub})$
    \State \hspace*{0.3in}{\bf If} {playing strong Stackelberg game}
    \State \hspace*{0.3in}$s_F^{*s_l} = \argmin_{s_F \in BR(s_L)} J_L(x(t),\{s_L(t),s_F(t)\}) $ 
    \State \hspace*{0.3in}{\bf Else If} {playing weak Stackelberg game}
      \State \hspace*{0.3in}$s_F^{*s_l} = \argmax_{s_F \in BR(s_L)} J_L(x(t),\{s_L(t),s_F(t)\}) $ 
    \State \hspace*{0.3in}{\bf End If}
    \State \hspace*{0.2in}{\bf End For}
    \State \hspace*{0.3in}$s_L^*(t)= \argmin_{s_L \in S_L} J_L(x(t),\{s_L(t),s_F^{*s_l}(t)\})$
    \State \hspace*{0.3in}$s_{F+}^*(t)= BR(s_L^*)$
    \State \hspace*{0.3in}$s^{*,Sh}(t)=\{s_L^*(t),s_{F+}^*(t)\}$

    \end{algorithmic}
\end{algorithm}

\begin{algorithm}[ht]

\caption{Stackelberg Pairwise Equilibrium }
\label{alg:SE_pairs}
\begin{algorithmic}
 
 \Input
    \Desc{$\mathcal{G}^S$}\hspace*{0.2in}{Stackelberg game to solve}
    \EndInput
    \Output
    \Desc{$s^{*,Sp}(t)$}\hspace*{0.2in}{Stackelberg pairwise equilibrium}
    
    \EndOutput
    \State {\bf Procedure}
    \end{algorithmic}
    \begin{algorithmic}[1]
    \setcounter{ALG@line}{0}
    
   \State \hspace*{0.2in} {$S^*_{ego}(t)=\{\}$}
\State \hspace*{0.2in}{\bf For} {$i = 2:N$} {\bf do}
\State \hspace*{0.3in}{$\mathcal{G}_i^S=\{\{1,i\},\mathcal{S}_1 \times \mathcal{S}_i,\{J_i\}_{i \in \{1,i\}}\}$}
\State \hspace*{0.3in} {$[s^*_{ego}(t),s_{i}^*(t)]=$ Algorithm \ref{alg:Strong Stackelberg Equilibrium} ($\mathcal{G}_i^S$)}
\State \hspace*{0.3in} {$S^*_{ego}(t)=S^*_{ego}(t)\cup \{s^*_{ego}(t)\}$}
\State \hspace*{0.2in}{\bf End For}

\State \hspace*{0.3in} {$s^*_1(t)=\text{MC}(S^*_{ego}(t))$}
\State \hspace*{0.3in} {$s^{*,Sp}(t) = \{s_1^*(t),s_2^*(t),...,s_N^*(t)\}$}

 \end{algorithmic}
 
\end{algorithm}


\begin{Remark}
\label{remark:SvW}
In autonomous driving applications, appropriate Stackelberg equilibrium (i.e., strong or weak) should be selected before using Stackelberg games in AV decision-making. We notice that if the ego vehicle is a follower in a scenario, then the strong Stackelberg equilibrium should always be employed, to enable considerate interactions with others. It is because in the weak Stackelberg equilibrium, the follower is designed to behave against the leader while defending its self-interest, which is not desirable since the AV is supposed to be considerate to others. On the other hand, if the ego vehicle is the leader, then both the strong and the weak Stackelberg equilibria make sense depending on the assumptions of other road users' intentions: If they are assumed to be `kind' and to behave in favor of others while keeping their self-interests optimized, then the strong Stackelberg equilibrium should be selected for a rationally `optimistic' ego vehicle; If otherwise, then the weak can be selected for a conservative ego vehicle. 
\end{Remark}

\section{Numerical Results}
\label{section: NumericalResults}
\begin{figure}
    \centering
    \resizebox{0.3\textwidth}{!}{\includegraphics{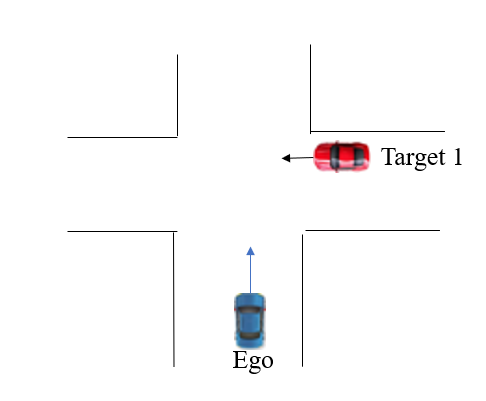}}
    \caption{2-Vehicle Unsupervised Intersection Scenario}
    \label{fig:unsupervised scenario}
\end{figure}

\begin{figure}
    \centering
    \resizebox{0.3\textwidth}{!}{\includegraphics{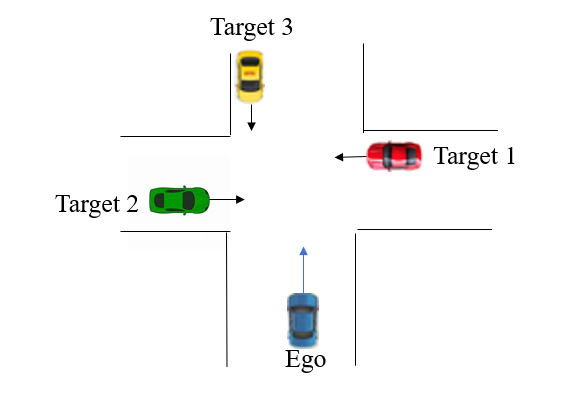}}
    \caption{4-Vehicle Unsupervised Intersection Scenario}
    \label{fig:unsupervised scenario 4players}
\end{figure}

In this section, we conduct numerical studies to test the performance of the games introduced in Sections \ref{section:Nash} and \ref{section:Stackelberg}. We test the ego vehicle performance, including safety (quantified by crash rate), travel efficiency (quantified by the ego vehicle average speed), and computational cost (quantified by the average running time of one decision-making process) against various target vehicle (i..e, the vehicle that is not under our control) strategies. In Section \ref{sub:2Player} we consider a two-vehicle intersection scenario as depicted in Figure \ref{fig:unsupervised scenario}. The performance of multi-vehicle scenarios is demonstrated in Section \ref{sub:4PlayerInt}, which is depicted in Figure \ref{fig:unsupervised scenario 4players}.

\subsection{Simulation Setup}

In the intersection-crossing scenario, we assume that all agents aim to cross the intersection safely and efficiently, i.e., tracking their own desired speeds while avoiding collisions. Motivated by this desire, we construct the cost functions  \eqref{eq:cost}. 
 \begin{equation}
     J_{i}(x(t),s(t))=\sum_{\tau=t}^{t+T-1 }Q_i(x(\tau),a(\tau)),
     \label{eq:cost}
 \end{equation}
where $Q_i(x(t),a(t))$ is described by Eq. \ref{eq:Q_i} and Eq. \ref{eq:Qij}.
\begin{equation}
    Q_i(x(t),a(t))=\frac{(v_i(t)-v_{i,d})^2}{v_{i,d}}+\sum_{j=1,j\neq i}^{N} Q_{ij}(x(t),a(t)),
    \label{eq:Q_i}
\end{equation}
\begin{dmath}
    {Q_{ij}(x(t),a(t)) = (\tanh(\beta (d_{x,c}^2-(x_i(t)-x_j(t))^2))+1)}\cdot\\{(\tanh(\beta (d_{y,c}^2-(y_i(t)-y_j(t))^2))+1)},
    \label{eq:Qij}
\end{dmath}
 where $v_{i,d}=10m/s$, $\forall i$, is vehicles' desired speed, $d_{x,c}=d_{y,c}=6m$ is a safe distance that is slightly larger than half the width of the road, and $\beta=1000$ is a parameter used to ensure the $\tanh$ function in Eq. \eqref{eq:Qij}  only takes the values of $-1$ or $1$. Additionally, player $i$'s position at time $t$ is denoted as $(x_i(t),y_i(t))$ and player $i$'s speed at time $t$ is denoted as $v_i(t)$. The action space of each player is $\{-2,-1,0,1,2\} m/s^2$. The prediction horizon used in Eq. \eqref{eq:cost} is $4s$. 

The vehicles' dynamics are described by the state space representation given in Eq. \eqref{eq:SSR} and Eq. \eqref{eq:SSR_output}. Let $\mathbf{x}_i(t)=[z_i(t) , v_i(t)]^T$ be the state of player $i$ at time $t$, where $z_i(t)$ is the position of player $i$ along that player's direction of travel and $v_i(t)$ is player $i$'s velocity. Let the control $u_i(t) = a_i(t)$ be the acceleration of player $i$. Then the vehicle dynamics of player $i$ are described by:

\begin{equation}
    \mathbf{x}_i(t+1)=\begin{bmatrix}
        1 & \Delta t  \\
        0 & 1
    \end{bmatrix} \mathbf{x}_i(t) + \begin{bmatrix}
        0   \\
        \Delta t
    \end{bmatrix} u_i(t) ,
    \label{eq:SSR}
\end{equation}
\begin{equation}
    \mathbf{y}_i(t)= \begin{bmatrix}
        1 & 0  \\
        0 & 1
    \end{bmatrix} \mathbf{x}_i(t) ,
    \label{eq:SSR_output}
\end{equation}
 where $\Delta t=0.5s$ is the sampling time.

To test the robustness of the ego vehicle strategies, we let the target vehicle take three different strategies:
\begin{itemize}
    \item ``Ideal'' meaning that the target vehicle also uses the outcome from the corresponding games, as the ego vehicle expected, representing the ideal situations;
    \item ``Simple rules'' meaning that the target vehicle brakes at its maximum allowed deceleration $(-2m/s^2)$ to a stop if it does not have the right of way (i.e., further away from the intersection compared to the ego vehicle) and is in danger of crashing (i.e., the distance between the target vehicle and any another vehicle is less than some distance $d_{s}$). Otherwise, the target vehicle chooses a strategy to reach or maintain its desired speed. Such a setting represents a safety- and rule- conscious vehicle, but its specific behaviors may not always be consistent with the ego vehicle expectation. 
    \item ``Constant speed'' meaning that the target vehicle simply keeps a constant speed and does not respond to the ego vehicle at all, representing a safety- and rule- agnostic vehicle. 
\end{itemize}


We let a strategy be composed of two actions, and each action lasts for half of the time horizon, i.e., $2s$. Since the strategy space in our study is relatively small, we evaluate each strategy to find the corresponding costs, and use the MATLAB function ``min" \cite{MATLAB_min} to find the optimal strategy. The runtime of simulations is measured by MATLAB ``tic-toc" functions \cite{MATLAB_Tic,MATLAB_Toc}, and the simulations are performed on a desktop with an Intel Core i9-12900k processor clocked at 3200 MHz and 64 GB of RAM. One hundred games of each of these behaviors were simulated in MATLAB. The ego vehicle initial conditions remain the same in all simulations, while the target vehicles' initial positions along the x-axis are randomized. 
Nash games are solved with the best response dynamics, and both strong and weak Stackelberg games are tested. 

For Stackleberg games, the ego always takes the role of leader, even in scenarios where people may assume another player is the leader based on road-priority rules. In the ideal behavior, it is not necessary to use any rules to determine the leader, since every player will know and agree to their given role. Additionally, based on how this paper defines the Stackelberg equilibrium, the follower can have a delay issue. The follower's optimal strategy according to definition \ref{def: SSE} (or \ref{def: WSE}) is the follower's best response to the leader's strategy. This requires knowledge about the leader's actual strategy, not what the follower thinks the leaders strategy should be. This could be achieved by having the follower start at time $t+1$ and react to the leader's strategy at time $t$. If the time step between times $t$ and $t+1$ is small, this setting would approximate simultaneous actions while still allowing the follower to react according to the definition of the Stackleberg equilibrium. By letting the ego always be the leader, we remove the need for such a complication. 

\subsection{2-Vehicle Intersection Crossing}
\label{sub:2Player}
\begin{table*}
\centering

\caption{Two Player Game}
\begin{tabular}{|c|ccc|ccc|ccc|}
\hline
Metric & \multicolumn{3}{c|}{Crashes per 100 Games} & \multicolumn{3}{c|}{Ave. Ego Speed (m/s)} & \multicolumn{3}{c|}{Ave. Decision Time (s)} \\ \hline
Behavior\textbackslash{}Game & \multicolumn{1}{c|}{SSE} & \multicolumn{1}{c|}{WSE} & NBR & \multicolumn{1}{c|}{SSE} & \multicolumn{1}{c|}{WSE} & NBR & \multicolumn{1}{c|}{SSE} & \multicolumn{1}{c|}{WSE} & NBR \\ \hline
Ideal & \multicolumn{1}{c|}{0} & \multicolumn{1}{c|}{0} & 0 & \multicolumn{1}{c|}{9.140} & \multicolumn{1}{c|}{9.140} & 9.164 & \multicolumn{1}{c|}{0.0053} & \multicolumn{1}{c|}{0.0052} & 0.0006 \\ \hline
Simple Rules & \multicolumn{1}{c|}{0} & \multicolumn{1}{c|}{0} & 0 & \multicolumn{1}{c|}{9.143} & \multicolumn{1}{c|}{9.143} & 9.143 & \multicolumn{1}{c|}{0.0053} & \multicolumn{1}{c|}{0.0053} & 0.0006 \\ \hline
Constant Speed & \multicolumn{1}{c|}{7} & \multicolumn{1}{c|}{7} & 4 & \multicolumn{1}{c|}{9.171} & \multicolumn{1}{c|}{9.171} & 9.142 & \multicolumn{1}{c|}{0.0053} & \multicolumn{1}{c|}{0.0053} & 0.0006 \\ \hline

\end{tabular}
\caption*{$D<5 m, v_1(t_0) = 4 m/s = v_2(t_0), v_d = 10 m/s$}
\label{table:5mcrash}
\end{table*}

In this subsection, we perform two comparative studies in a 2-vehicle unsupervised intersection crossing: 1) Nash equilibrium vs Stackleberg equilibrium, and 2) Strong vs Weak Stackelberg equilibrium.  

\begin{Study}[Statistical Comparison of Nash Game and Stackleberg Game]\label{Study:NGvsSG}
We compare the crash rate, average speed of the ego vehicle, and the average computational time needed to calculate the optimal strategy (which we call decision time). The crash rate is the number of scenarios where the ego comes within $5m$ of the target vehicle out of the total 100 scenarios. The desired speed of each vehicle is $10m/s$, and the ego vehicle's travel efficiency is determined by the difference between its desired and average speeds. From Table \ref{table:5mcrash} we make the following observations:
\begin{itemize}
    \item In terms of safety, both games are only effective in ensuring the ego vehicle's safety if the target is safety-conscious. If the target is not safety-conscious, the Nash equilibrium is shown to be slightly safer than either Stackelberg equilibrium in the tested scenarios.
    \item In terms of travel efficiency, both games have similar performance in the conducted tests. 
    \item In terms of computational cost, both games lead to similar runtime in this two-player case, and the decision time is affordable (less than $0.01s$ on average).
\end{itemize}

While the ideal and simple-rules behaviors both are shown to be safe, we note that the simple-rules behavior may not be able to safely navigate more complex scenarios, which is confirmed in the next subsection in 4-vehicle scenarios. 

\end{Study}
\begin{Study}[Comparison of Stackelberg Equilibriums]\label{Study:TwoStacks}
    The strong and weak Stackelberg games have similar performance in the statistical comparison. 
    It is because the utility function given by Eq. \eqref{eq:cost} makes it unlikely that a player have a tie between two of their strategies. To better observe their difference, we adjust the game settings in this study.  Specifically, we restrict the actions of a player to the set $\mathcal{A}_i = \{-1,0,1\}m/s^2$ and strategies are restricted to only consist of one action lasting for the whole horizon. The cost matrix of the leader and the follower are designed in Table \ref{table: sse v wse}. 
    \begin{table}[ht!]
\begin{tabular}{|c|c|c|c|}
\hline
Leader\textbackslash{}Follower & -1     & 0       & 1       \\ \hline
-1                             & (5,10) & (5,5)   & (5,0)   \\ \hline
0                              & (0,10) & (0,5)   & (5,5)   \\ \hline
1                              & (5,10) & (10,10) & (15,10) \\ \hline
\end{tabular}
\caption{Cost matrix to compare strong and weak Stackelberg games}
\label{table: sse v wse}
\end{table}

With such a cost matrix, the strong Stackelberg equilibrium is the strategy $\{0,0\}m/s^2$. Both of the strategies $\{0,1\}m/s^2$ and $\{-1,1\}m/s^2$ are weak Stackelberg equilibriums, indicating that the leader is more, or at most the same, conservative in the weak compared to that in the strong. 
    This is also observed from the simulation. Figure \ref{fig: two Stackelberg algorithms} shows snapshots of this simulation at key times. By comparing the ego vehicle positions and velocities at $t=3.5s$, we can observe that the ego vehicle is more conservative in the weak Stackelberg game than in the strong. It is because the ego expects a more aggressive action from the target and reacts accordingly. 
    

\begin{figure*}[ht!]
\centering
    \begin{subfigure}[b]{\textwidth}
    \centering
    \resizebox{\textwidth}{!}{\includegraphics{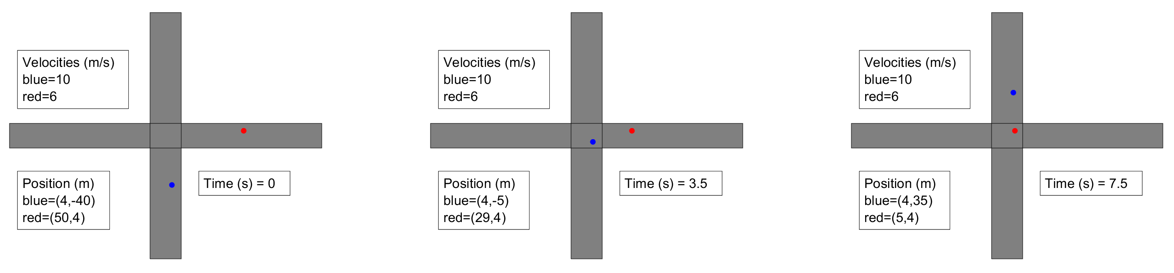}}
    \caption{Strong Stackelberg equilibrium}
    \label{fig:sse_example}
    \end{subfigure}
    \begin{subfigure}[b]{\textwidth}
    \centering
    \resizebox{\textwidth}{!}{\includegraphics{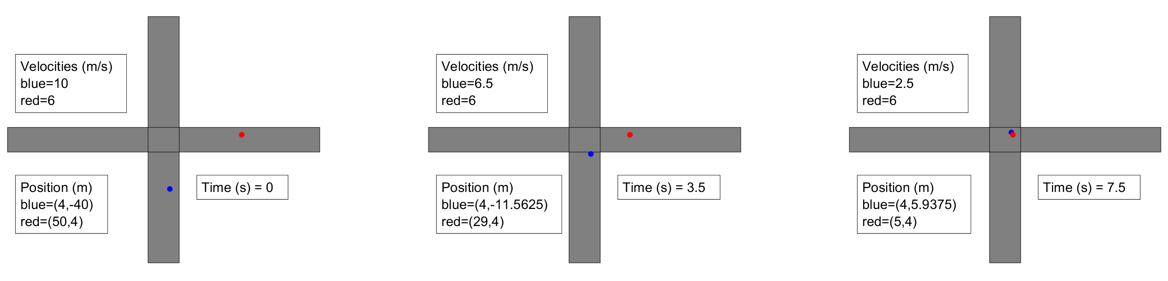}}
    \caption{Weak Stackelberg equilibrium}
    \label{fig:wse_example}
    \end{subfigure}
    \caption{Different Stackelberg Equilibriums}
    \label{fig: two Stackelberg algorithms}
\end{figure*}
\end{Study}

\subsection{4-vehicle Intersection Crossing}
\label{sub:4PlayerInt}
In this subsection, we perform a comparative study similar to Section  \ref{sub:2Player} in 4-vehicle scenarios. Table \ref{table:5mcrash_4players} shows the results of solving the 4-player game and Table \ref{table:5mcrash_4players_pairwise} shows the results of solving pairwise games. Table \ref{table:5mcrash_4players} leads to the following observations:

\begin{itemize}
    \item In terms of safety, if the target vehicles  behave ideally, then both games are effective in ensuring the ego vehicle safety. If not, then the Stackelberg equilibrium is  slightly safer than the Nash equilibrium in the tested scenarios.
    \item In terms of travel efficiency, both games perform similarly.   
    \item In terms of computational cost, Stackelberg games are much more computationally expensive compared to Nash games ($3s$ vs. $0.002s$).
\end{itemize}

By comparing the performance of 4-player games and pairwise games (i.e., Tables \ref{table:5mcrash_4players} and \ref{table:5mcrash_4players_pairwise}), we have the following observations:
\begin{itemize}
    \item In terms of safety, 4-player game is much more effective in keeping the ego vehicle safe compared to pairwise games, indicating better robustness. 
    \item In terms of travel efficiency, the two classes of games perform similarly. 
    \item In terms of computational complexity, pairwise games are more computationally efficient, especially for Stackelberg games.
\end{itemize}

\begin{table*}
\centering
 \caption{Four Player Game}
\begin{tabular}{|c|ccc|ccc|ccc|}
\hline
Metric & \multicolumn{3}{c|}{Crashes per 100 Games} & \multicolumn{3}{c|}{Ave. Ego Speed (m/s)} & \multicolumn{3}{c|}{Ave. Decision Time (s)} \\ \hline
Behavior\textbackslash{}Game & \multicolumn{1}{c|}{SSE} & \multicolumn{1}{c|}{WSE} & NBR & \multicolumn{1}{c|}{SSE} & \multicolumn{1}{c|}{WSE} & NBR & \multicolumn{1}{c|}{SSE} & \multicolumn{1}{c|}{WSE} & NBR \\ \hline
Ideal & \multicolumn{1}{c|}{0} & \multicolumn{1}{c|}{0} & 0 & \multicolumn{1}{c|}{9.139} & \multicolumn{1}{c|}{9.139} & 9.195 & \multicolumn{1}{c|}{3.2992} & \multicolumn{1}{c|}{3.2981} & 0.0015 \\ \hline
Simple Rules & \multicolumn{1}{c|}{1} & \multicolumn{1}{c|}{1} & 3 & \multicolumn{1}{c|}{8.989} & \multicolumn{1}{c|}{8.989} & 9.154 & \multicolumn{1}{c|}{3.2905} & \multicolumn{1}{c|}{3.2720} & 0.0015 \\ \hline
Constant Speed & \multicolumn{1}{c|}{2} & \multicolumn{1}{c|}{2} & 2 & \multicolumn{1}{c|}{9.197} & \multicolumn{1}{c|}{9.197} & 9.148 & \multicolumn{1}{c|}{3.2614} & \multicolumn{1}{c|}{3.2618} & 0.0015 \\ \hline
\end{tabular}

\caption*{$D<5m, v_i(t_0) = 4 m/s \: \forall i \in \mathcal{N}, v_d = 10 m/s$}
\label{table:5mcrash_4players}
\end{table*}

\begin{table*}
\centering
\caption{Four Players, Pairwise Games}
\begin{tabular}{|c|ccc|ccc|ccc|}
\hline
Metric & \multicolumn{3}{c|}{Crashes per 100 Games} & \multicolumn{3}{c|}{Ave. Ego Speed (m/s)} & \multicolumn{3}{c|}{Ave. Decision Time (s)} \\ \hline
Behavior\textbackslash{}Game & \multicolumn{1}{c|}{SSE} & \multicolumn{1}{c|}{WSE} & NBR & \multicolumn{1}{c|}{SSE} & \multicolumn{1}{c|}{WSE} & NBR & \multicolumn{1}{c|}{SSE} & \multicolumn{1}{c|}{WSE} & NBR \\ \hline
Ideal & \multicolumn{1}{c|}{0} & \multicolumn{1}{c|}{0} & 0 & \multicolumn{1}{c|}{9.140} & \multicolumn{1}{c|}{9.140} & 9.122 & \multicolumn{1}{c|}{0.0158} & \multicolumn{1}{c|}{0.0158} & 0.0017 \\ \hline
Simple Rules & \multicolumn{1}{c|}{5} & \multicolumn{1}{c|}{5} & 5 & \multicolumn{1}{c|}{9.136} & \multicolumn{1}{c|}{9.136} & 9.048 & \multicolumn{1}{c|}{0.0158} & \multicolumn{1}{c|}{0.0158} & 0.0017 \\ \hline
Constant Speed & \multicolumn{1}{c|}{22} & \multicolumn{1}{c|}{22} & 4 & \multicolumn{1}{c|}{9.133} & \multicolumn{1}{c|}{9.133} & 9.009 & \multicolumn{1}{c|}{0.0157} & \multicolumn{1}{c|}{0.0161} & 0.0017 \\ \hline

\end{tabular}
\caption*{$D<5m, v_i(t_0) = 4 m/s \: \forall i \in \mathcal{N}, v_d = 10 m/s$}
\label{table:5mcrash_4players_pairwise}
\end{table*}

\section{Conclusion}
\label{section:conclusion}

This paper tested the safety, travel efficiency, and computational costs of utilizing Nash and Stackelberg games in the AV decision-making. The results show that if the surrounding agents behave ideally, then both games can ensure the ego vehicle safety. However, in non-ideal situations, the Nash game performs better than the Stackelberg game in most situations. 
In addition, solving the Nash game is shown to be more computationally efficient than solving the Stackelberg game when multiple agents get involved. Although pairwise games can address this scalability challenge, they often lead to less desirable performance in terms of safety and robustness. 



\bibliographystyle{IEEEtran}
\bibliography{IEEEabrv,references}

\end{document}